\documentclass[%
reprint,
superscriptaddress,
 amsmath,amssymb,
 aps,
prb,
]{revtex4-1}

\usepackage{graphicx}
\usepackage{dcolumn}
\usepackage{bm}
\usepackage{hyperref}
\usepackage{sidecap}

\begin{document}
\title{Magnetic domain tuning and the emergence of bubble domains in the bilayer manganite La$_{2-2x}$Sr$_{1+2x}$Mn$_2$O$_7$ ($x=0.32$)}

\author{Juyoung Jeong}
\affiliation{Center for Artificial Low Dimensional Electronic Systems, Institute for Basic Science, 77 Cheongam-Ro, Nam-Gu, Pohang 790-784, Korea}
\affiliation{Department of Physics, Pohang University of Science and Technology, Pohang 790-784, Korea}

\author{Ilkyu Yang}
\affiliation{Center for Artificial Low Dimensional Electronic Systems, Institute for Basic Science, 77 Cheongam-Ro, Nam-Gu, Pohang 790-784, Korea}
\affiliation{Department of Physics, Pohang University of Science and Technology, Pohang 790-784, Korea}

\author{Jinho Yang}
\affiliation{Center for Artificial Low Dimensional Electronic Systems, Institute for Basic Science, 77 Cheongam-Ro, Nam-Gu, Pohang 790-784, Korea}
\affiliation{Department of Physics, Pohang University of Science and Technology, Pohang 790-784, Korea}

\author{Oscar E. Ayala-Valenzuela}
\affiliation{Center for Artificial Low Dimensional Electronic Systems, Institute for Basic Science, 77 Cheongam-Ro, Nam-Gu, Pohang 790-784, Korea}
\affiliation{Department of Physics, Pohang University of Science and Technology, Pohang 790-784, Korea}

\author{Dirk Wulferding}
\affiliation{Center for Artificial Low Dimensional Electronic Systems, Institute for Basic Science, 77 Cheongam-Ro, Nam-Gu, Pohang 790-784, Korea}
\affiliation{Department of Physics, Pohang University of Science and Technology, Pohang 790-784, Korea}

\author{J.-S. Zhou}
\affiliation{Texas Materials Institute, University of Texas, Austin, Texas 78712, USA}

\author{John B. Goodenough}
\affiliation{Texas Materials Institute, University of Texas, Austin, Texas 78712, USA}

\author{Alex de Lozanne}
\affiliation{Department of Physics, University of Texas, Austin, Texas 78712, USA}

\author{John F. Mitchell}
\affiliation{Materials Science Division, Argonne National Laboratory, Argonne, Illinois 60439, USA}

\author{Neliza Leon}
\affiliation{MPA-CMMS, Los Alamos National Laboratory, Los Alamos, New Mexico 87545, USA}

\author{Roman Movshovich}
\affiliation{MPA-CMMS, Los Alamos National Laboratory, Los Alamos, New Mexico 87545, USA}

\author{Yoon Hee Jeong}
\affiliation{Department of Physics, Pohang University of Science and Technology, Pohang 790-784, Korea}

\author{Han Woong Yeom}
\affiliation{Center for Artificial Low Dimensional Electronic Systems, Institute for Basic Science, 77 Cheongam-Ro, Nam-Gu, Pohang 790-784, Korea}
\affiliation{Department of Physics, Pohang University of Science and Technology, Pohang 790-784, Korea}

\author{Jeehoon Kim}
\email[]{Corresponding author: jeehoon@postech.ac.kr}
\affiliation{Center for Artificial Low Dimensional Electronic Systems, Institute for Basic Science, 77 Cheongam-Ro, Nam-Gu, Pohang 790-784, Korea}
\affiliation{Department of Physics, Pohang University of Science and Technology, Pohang 790-784, Korea}

\date{\today}

\begin{abstract}

We report a magnetic force microscopy study of the magnetic domain evolution in the layered manganite La$_{2-2x}$Sr$_{1+2x}$Mn$_2$O$_7$ (with $x=0.32$). This strongly correlated electron compound is known to exhibit a wide range of magnetic phases, including a recently uncovered biskyrmion phase. We observe a continuous transition from dendritic to stripe-like domains, followed by the formation of magnetic bubbles due to a field- and temperature dependent competition between in-plane and out-of-plane spin alignments. The magnetic bubble phase appears at comparable field- and temperature ranges as the biskyrmion phase, suggesting a close relation between both phases. Based on our real-space images we construct a temperature-field phase diagram for this composition.

\begin{description}

\item[PACS numbers]{68.37.Rt, 75.60.Ch, 75.70.Kw}

\end{description}
\end{abstract}

\maketitle

\section{Introduction}

Perovskite manganites with strongly coupled charge, spin, and lattice degrees of freedom provide a treasure chest of interesting and exotic physics.~\cite{salamon-01, haghiri-03} One of the most prominent phenomena studied in manganites is the colossal magnetoresistance (CMR).~\cite{ling-00} In addition, a plethora of magnetic phases has been uncovered in the layered manganite La$_{2-2x}$Sr$_{1+2x}$Mn$_2$O$_7$, corresponding to the $n=2$ member of the Ruddlesden-Popper series ($R,A$)$_{n+1}$Mn$_n$O$_{3n+1}$ ($R$ and $A$ being trivalent rare earth and divalent alkaline earth ions, respectively), such as antiferromagnetic metallic, ferromagnetic metallic, canted antiferromagnetic, insulating antiferromagnetic, and charge ordered.~\cite{ling-00} The transitions among these states can be controlled by the doping level and are amplified in comparison to members of the $n=3$ family due to the low dimensionality of the system, leading to a huge CMR effect. In particular, in the ferromagnetic phase, the easy axis can be tuned from out-of-plane (for $0.30 \leq x < 0.32$) to in-plane (for $0.32 < x \leq 0.40$).~\cite{kimura-00, kubota-99} Hence, the samples with $x \approx 0.32$ (``LSMO-032'') are close to an instability and a spin reorientation transition (SRT) can be induced easily by changing temperature and magnetic field. Several studies have been carried out to detail the magnetic structure of LSMO close to the instability. These include low-field magnetic force microscopy (MFM) up to 240 G,~\cite{huang-08} zero-field Lorentz transmission electron microscopy (TEM),~\cite{asaka-05} magneto-optical imaging,~\cite{welp-00} scanning tunneling microscopy (STM),~\cite{kim-13} and scanning Hall probe microscopy studies.~\cite{fukumura-99} Intriguingly, a recent Lorentz TEM study revealed the formation of a biskyrmion lattice in thin LSMO samples under an applied magnetic field of about 0.3 T.~\cite{yu-14} However, until now there is no comprehensive report about the development of magnetic domains in a bulk single crystal of LSMO-032 exceeding magnetic fields of 240 G.~\cite{bryant-15}

Here, we report a detailed investigation of the domain evolution as a function of both temperature and magnetic field in bulk LSMO-032 via MFM. We trace the SRT and find the formation of bubble domains in the vicinity of the SRT. In contrast to previous studies we directly observe the transition from dendritic to bubble domains as a function of the magnetic field. We also study the reversibility and unveil a weak hysteretic behavior. Our temperature dependent MFM investigation evidences clearly the transition between bubbles and dendritic domains. The MFM images reveal the gradual character of the SRT. We construct a magnetic field - temperature (\textbf{H}-\textit{T}) phase diagram based on MFM images obtained in applied magnetic fields of 0 T $\leq$ $\mu_0$\textbf{H} $\leq$ 0.4 T and in a temperature range of 13.5 K $\leq$ $T$ $\leq$ 70 K. Our study serves as a counterpart to the recent investigation of biskyrmions in a thin LSMO specimen~\cite{yu-14} and highlights the role of shape anisotropy in the formation of skyrmion phases in centrosymmetric compounds.

\section{Experimental Details}

Single crystals of La$_{1.36}$Sr$_{1.64}$Mn$_2$O$_7$ were grown by the traveling-floating-zone method in an image furnace.~\cite{zhou-00} Magnetization measurements were performed in a dc superconducting quantum interference device (Quantum Design). The dimensions of the sample under investigation are 0.87(2) mm $\times$ 0.67(2) mm $\times$ 0.40(2) mm. The single crystal was oriented via a four-circle diffractometer / goniometer.

MFM measurements were performed in a low temperature MFM system with a home-built MFM probe inside a cryostat with a superconducting magnet in a field- and temperature range of 0 -- 7 T and 4 -- 300 K, respectively.~\cite{jinho-mfm} All experiments were carried out with commercially available tips (PPP-MFMR, Nanosensors). In MFM measurements, the magnetic force generated by magnetic domains is detected quantitatively in terms of the resonant frequency shift of the cantilever: a repulsive (attractive) magnetic force results in a positive (negative) frequency shift.

\section{Experimental Results and Discussion}

\begin{figure}
\label{figure1}
\centering
\includegraphics[width=8cm]{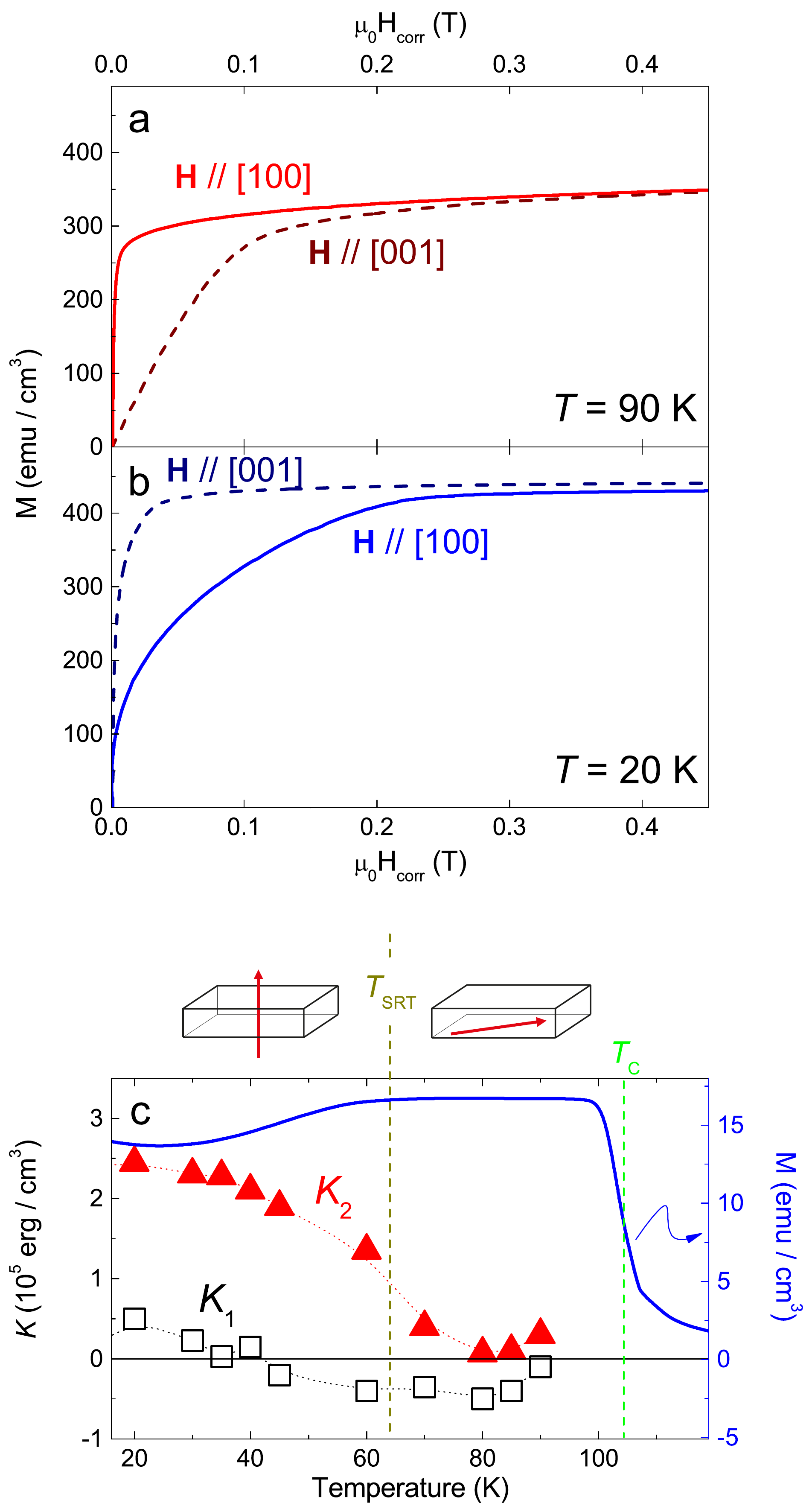}
\caption{(Color online) Magnetization curves at (a) 90 K and (b) 20 K measured parallel (dashed lines) and perpendicular (solid lines) to the $c$ axis. (c) Temperature dependence of the anisotropy constants $K_1$ (empty black squares) and $K_2$ (solid red triangles). Solid blue line: temperature dependent magnetization curve, with $\mu_0 H = 5$ mT $\perp c$ axis.}
\end{figure}

We will first focus on the global magnetization results before discussing our MFM images. Magnetization curves obtained at two different temperatures are shown in Figs. 1(a) ($T = 90$ K) and (b) ($T = 20$ K) with magnetic fields applied along the [100] direction (solid curves) and [001] direction (dashed curves). The saturation magnetization at $T = 20$ K and $T = 90$ K is 440 emu/cm$^3$ and 360 emu/cm$^3$, respectively. No significant differences are found in curves between \textbf{H} // [100] and \textbf{H} // [110] (not shown), which can be expected considering only a relatively small in-plane anisotropy.~\cite{welp-01} In addition, we do not find any evidence for hysteretic behavior in any field direction in the \textit{M}-\textbf{H} curve. In order to directly compare the different magnetization directions, we use the corrected magnetic field $\textbf{H}_{corr}$ as the $x$-axis, which is obtained by subtracting the demagnetizing field $\textbf{H}_d = D \cdot \textbf{M}$ from the applied field \textbf{H}. The demagnetizing factors $D$ are based on the sample's dimension. We find $D = 0.22$ for [100] and $D = 0.49$ for [001].~\cite{aharoni-98} From the magnetization curves we find an in-plane easy axis at 90 K and an easy axis along the crystallographic $c$ axis at 20 K. Temperature dependent magnetization measurements in a weak applied field of $\mu_0$\textbf{H} = 5 mT yield the Curie temperature $T_C$ = 105 K and a SRT from in-plane to out-of-plane at $T_{SRT} = 63$ K [see blue curve in Fig. 1(c)]. The absolute value of $T_C$ together with the sharpness of the transition agrees well with previous reports on LSMO-032 and underlines the sample quality in terms of doping homogeneity.~\cite{medarde-99} The change in easy axis in samples with $x \approx 0.32$ originates from an additional contribution of the dipolar interaction (favoring an in-plane spin alignment) among Mn moments to the magnetocrystalline anisotropy constants $K_1$ and $K_2$ (favoring an out-of-plane spin alignment for positive values).~\cite{welp-00, welp-01} To gain a deeper insight into the SRT process, we calculate the first and second order anisotropy constants ($K_1$ and $K_2$, respectively) from a fit applied to the hard-axis \textit{M}-\textbf{H} curves at various temperatures (see Fig. 1(c) and Ref.~\cite{durst-86} for details). For $K_1 > 0$ and $K_2 > 0$, the easy axis is along the $c$ axis. However, for $K_1 < 0$ and $K_2 < -K_1 /2$ the easy axis is within the $ab$ plane. For $K_1 < 0$ and $K_2 > -K_1 /2$ the easy axis is along a cone centered around the $c$ axis with an opening angle of $\theta = \mathrm{sin}^{-1} \left( \sqrt{\frac{\vert K_1 \vert}{2K_2}} \right)$.~\cite{coey-10} In our case, above 80 K the easy axis is in-plane while below 40 K the easy axis is out-of-plane. The temperature evolution of the anisotropy is related to a change in orbital occupancy from $d_{3z^2-r^2}$ at low temperatures to predominantly $d_{x^2-y^2}$ at high temperatures.~\cite{hirota-02} The small, positive $K_1$ values below 40 K can be associated with a weak uniaxial anisotropy. Interestingly, a previously reported temperature dependence study of $K_1$ in LSMO-032 found a strong increase with lowering temperature, similar to our $K_2$ temperature behavior.~\cite{welp-00} The fact that we only observe a weak, gradual increase of $K_1$ as we decrease the temperature underlines the delicate balance between out-of-plane and in-plane spin alignment in our sample, as well as the sensitivity of LSMO-032 to tiny compositional differences.

\begin{figure*}
\label{figure2}
\centering
\includegraphics[width=17.5cm]{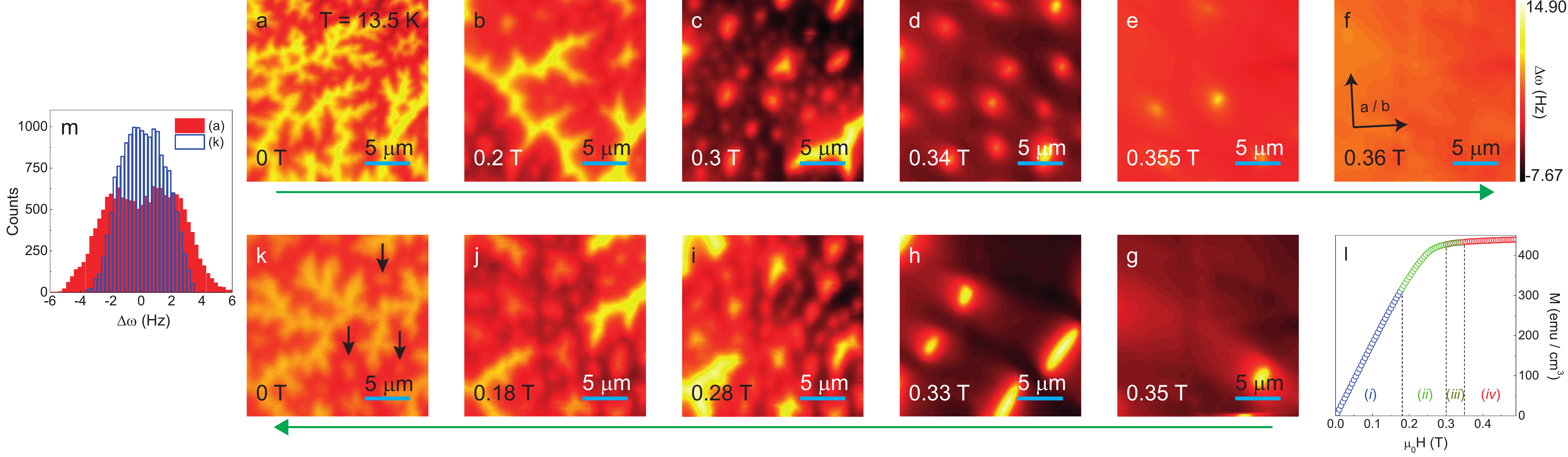}
\caption{(Color online) (a) -- (k) Magnetic field evolution of the domain structures at $T = 13.5$ K. Magnetic fields are indicated in the figures. The scale bar corresponds to 5 $\mu$m. The crystallographic orientation for all images is indicated in (f). (l) Magnetic hysteresis loop at $T = 13.5$ K with \textbf{H} // $c$ axis. The following phases are indicated: $(i)$ dendritic domains, $(ii)$ irregular stripes and bubbles, $(iii)$ isolated bubbles, and $(iv)$ saturation. (m) Histograms of the 0 T MFM images before (solid red) and after (empty blue) field cycling.}
\end{figure*}

In order to elucidate the magnetic domain formation of the LSMO-032 single crystal in an applied magnetic field, we image magnetic domain structures at $T = 13.5$ K after zero-field cooling (ZFC) through $T_C$. We start from the thermally demagnetized state and apply a stepwise increasing / decreasing magnetic field in the range of 0 -- 0.36 T. A selection of these images is shown in Figs. 2(a) -- (k). The green arrows denote the measurement sequence. The domains at zero-field exhibit clear dendritic structures with a nearly random orientation (a weak orientation preference is observed, running from the lower left to the upper right corner), signaling only a weak in-plane anisotropy directed to 45$^{\circ}$ from the crystallographic $a$ and $b$ axes [see Fig. 2(a)]. This is in accordance with a previous magnetization study, where a minor in-plane anisotropy along the [110] direction was revealed.~\cite{welp-01} These patterns result from an energy minimizing process that compromises between energy reduction from an overall decrease in the surface magnetic-pole energy and the concomitant energy increase from the creation of new domain walls.~\cite{huang-08} Similar domain patterns have been observed in samples with a uniaxial anisotropy.~\cite{unguris-89} Dendritic domains remain at magnetic fields up to 0.18 T. For higher fields the magnetic landscape displays an inhomogeneous nucleation of bubble domains (with a magnetization opposite to the external field) in coexistence with irregular stripe domains [see bright spots in Figs. 2(b) and (c)]. A crossover from the coexistence phase to isolated large bubbles (diameter $\approx 5$ $\mu$m) occurs upon increasing the magnetic field from 0.3 T to 0.355 T [see Figs. 2(c) -- (e)]. At $\mu_0 H > 0.355$ T, a transition to a uniformly magnetized phase takes place [Fig. 2(f)]. This is supported by the magnetization curve obtained at $T = 13.5$ K and with \textbf{H} // $c$ axis [Fig. 2(l)], where a saturation is reached around 0.35 T. Upon decreasing the magnetic field again, we observe the reappearance of bubbles around 0.35 T followed by an inhomogeneous but reversible transition from bubble to stripe domains around 0.18 T [see Figs. 2(g) -- (j)]. Note, that the direction of the weak in-plane anisotropy is preserved, as evidenced, e.g., in Fig. 2(h). Decreasing the magnetic field further leads to the return of dendritic structures. Here, we point out that although qualitatively the dendritic structures in Fig. 2(a) and (k) are the same, the local patterns are different.  A similar behavior is observed at magnetic fields around 0.33 -- 0.35 T, where bubble domains vanish (upon increasing fields) and nucleate (upon decreasing fields) at different locations. These observations highlight the absence of chemical gradients but a homogeneous stoichiometry throughout the sample, suggesting that the domain structure and the nucleation sites are not related to local defects. On the other hand, some features that can be associated with metastable bubbles remain at zero fields [see arrows in Fig. 2(k)]. In Fig. 2(m) histograms of the 0 T MFM images before (solid red bars) and after (empty blue bars) the field-cycling are compared. Both histograms show a similar distribution of negative and positive $\Delta \omega$ values centered around $\Delta \omega = 0$, which agrees very well with the absence of any hysteretic behavior in the $M$-\textbf{H} curve. At the same time, the distribution is considerably narrower after the field-cooling, which can be directly observed in the reduced contrast of Fig. 2(k) with respect to Fig. 2(a). We can therefore attest a reduction in the magnitude of the out-of-plane magnetic moment and hence a small magnetic field hysteresis in LSMO-032. This was not picked up in previous magnetization measurements and remains an open issue.~\cite{welp-00} The same sequence of images was also taken at $T = 50$ K (not shown), where bubble structures appeared in the same magnetic field range. Regardless of the temperature and the magnetic field, no long-range ordered bubble-lattice structures have been observed. This is contrasted by the Lorentz TEM study, where a well-arranged lattice of biskyrmions has been reported in thin LSMO specimens with $x=0.315$.~\cite{yu-14}

\begin{SCfigure*}
\label{figure3}
\centering
\includegraphics[width=12.5cm]{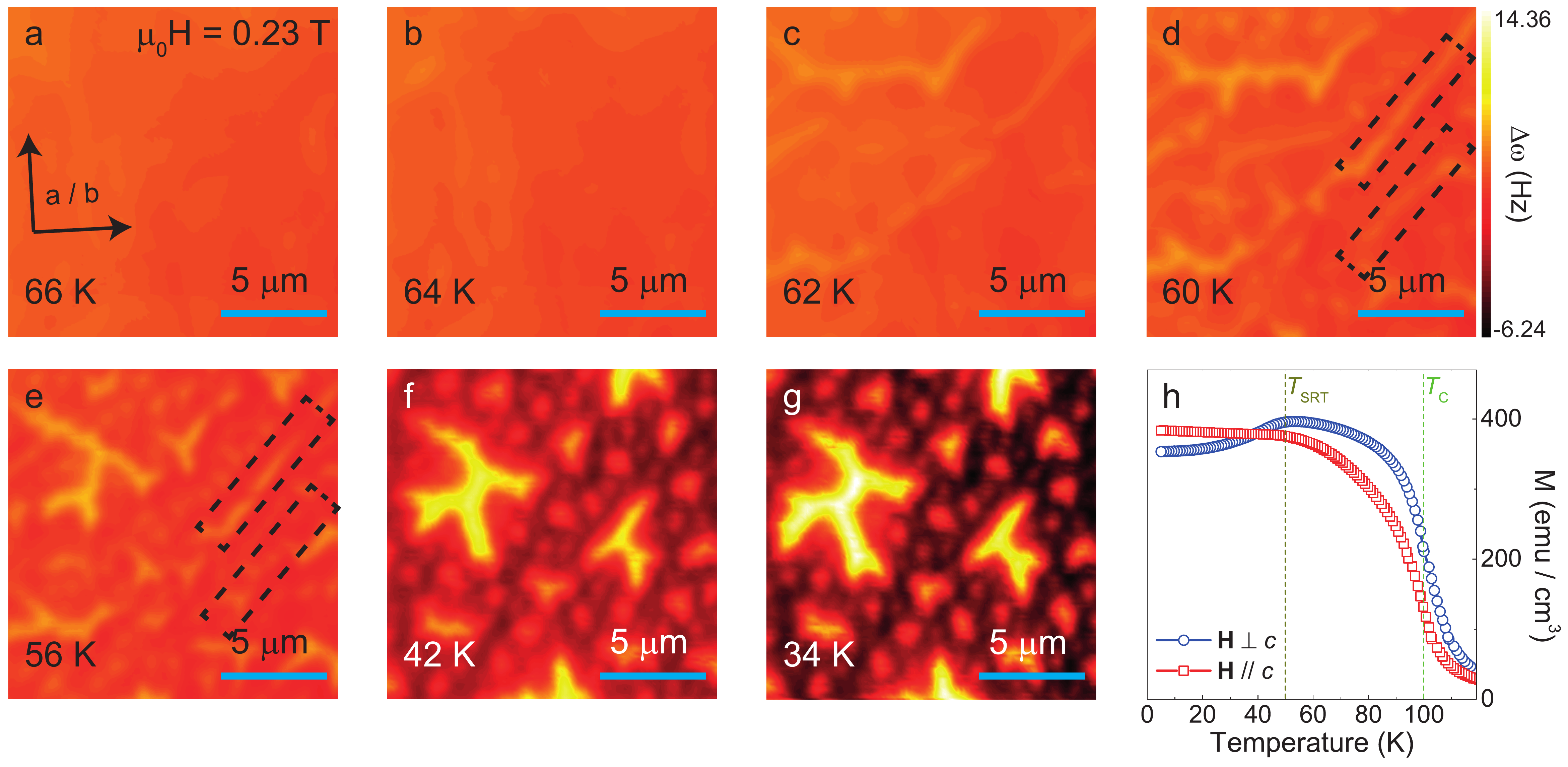}
\caption{(Color online) (a) -- (g) Evolution of magnetic domain structures in LSMO-032 with temperatures ranging from 66 K to 34 K and $\mu_0 H = 0.23$ T along the $c$ axis. The scale bar corresponds to 5 $\mu$m. The crystallographic orientation for all images is indicated in (a). (h) Temperature dependence of the magnetization with $\mu_0 H = 0.23$ T applied perpendicular (blue circles) and parallel (red squares) to the $c$ axis.}
\end{SCfigure*}

In the following, we investigate the temperature dependence of the magnetic domain structures. The sample was field-cooled through $T_C$ with a fixed magnetic field of $\mu_0 H = 0.23$ T, which left the sample with the coexistence of stripes and bubbles below $T_{SRT}$. The temperature dependence of the domain structure of LSMO-032 in zero- and weak magnetic fields ($\mu_0 H = 24$ mT) has been investigated previously via MFM, without observing any signs of bubble domains.~\cite{huang-08} Instead, a change from in-plane (with clear domain walls) to dendritic structures around $T_{SRT} = 85$ K was observed. This is considerably higher than our $T_{SRT}$ of about 63 K at zero fields and further highlights the delicate dependence of the sample's magnetic properties on minute differences in composition around $x=0.32$. The dendritic domains show drastic changes between 63 K and 45 K.~\cite{asaka-05, welp-99} This evolution can be associated with a ripple state that evolves linearly during the spin reorientation.~\cite{asaka-05} The ripple state is related to the layered crystal structure that enables dipolar interactions between Mn-ions (in-plane) and the orbital transition.~\cite{asaka-05, kimura-98} Figures 3(a) -- (g) show the evolution of the magnetic domains upon cooling the sample from $T = 66$ to 34 K, and Fig. 3(h) plots the temperature dependence of the magnetization with \textbf{H} // [001] as well as $\perp$ [001], and $\mu_0 H = 0.23$ T. At this field, $T_{SRT}$ is reduced from $\sim 63$ K at $\mu_0 H = 5$ mT [see Fig. 1(c), blue line] to $\sim 50$ K. On the other hand, our MFM images reveal that the spin reorientation process is rather gradual and occurs over a wide temperature range. For temperatures between 80 K and 64 K, the sample hosts in-plane domains~\cite{huang-08} or conically aligned spins between the $ab$ plane and the $c$ axis,~\cite{welp-00} hence producing MFM images of weak contrast. Weak domain structures emerge around 62 K, and from 56 K to 42 K the contrast increases dramatically, signaling an increase in out-of-plane moments. Around 60 K, stripe domains and bubbles begin to coexist [see Fig. 3(d) -- (g)]. Here, two regimes are observed: First, between 60 K and 45 K [see Fig. 3(e)] disordered structures coexist with stripes and bubble domains. Subsequently, below 45 K [see Figs. 3(f) -- (g)], the images show disordered bubble structures with features similar to those observed at 13.5 K at comparable magnetic fields. As evidenced in Fig. 2, a coexistence of stripes and bubbles even remains at intermediate fields ($\sim$0.2 -- 0.3 T) down to low temperatures. Interestingly, the nucleation of bubbles segmenting stripe domains can be directly observed [see black dashed boxes in Figs. 3(d) and (e)].

\begin{figure}
\label{figure4}
\centering
\includegraphics[width=8cm]{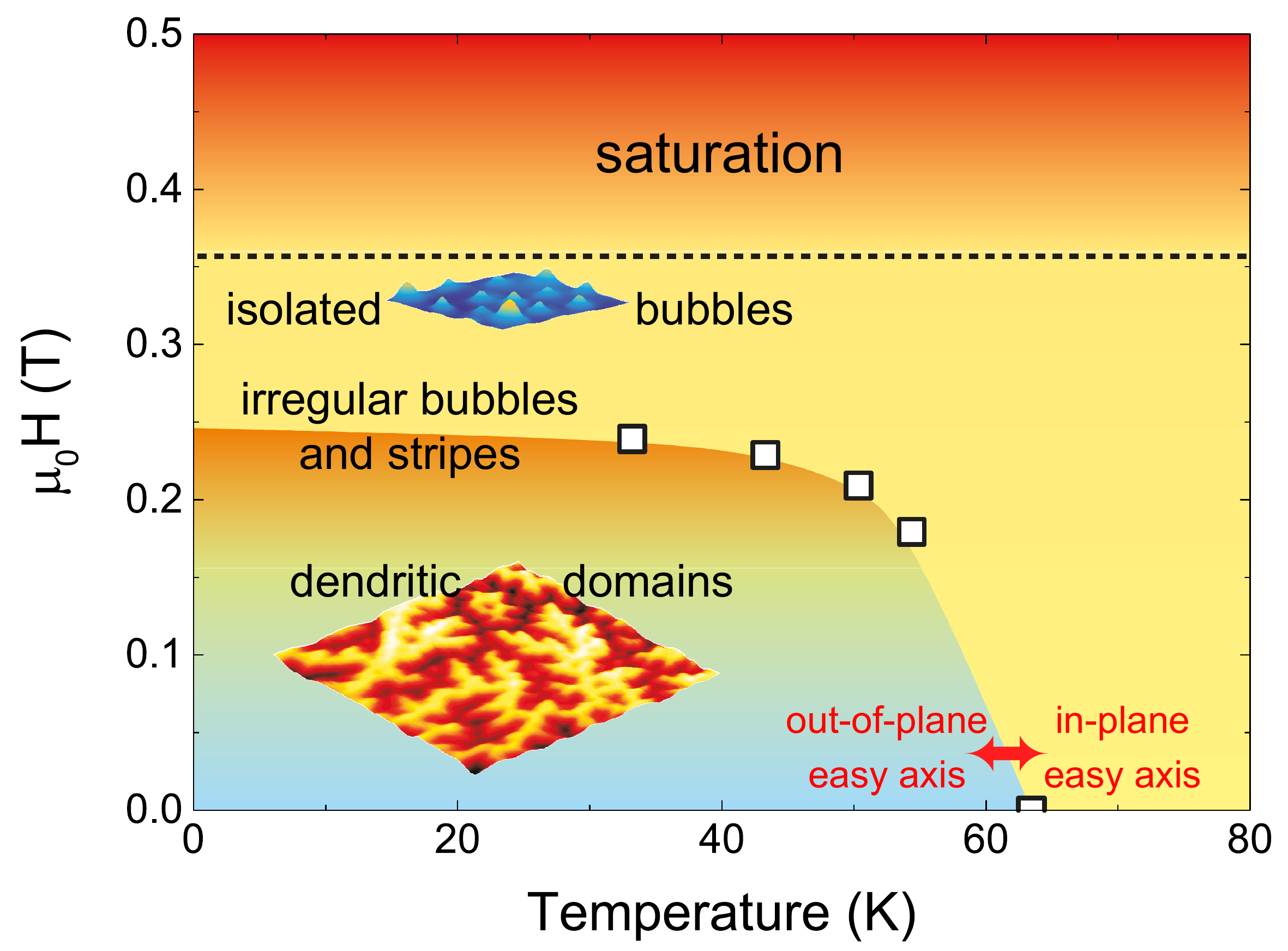}
\caption{(Color online) Schematic phase diagram of a LSMO-032 single crystal with the magnetic field applied along the $c$ axis. The white squares correspond to the SRT obtained from magnetization curves. The geometrical demagnetization factor along this direction is $D \approx 0.49$.}
\end{figure}

In Fig. 4 we summarize our MFM and magnetization results in a schematic field -- temperature (\textbf{H}-\textit{T}) phase diagram of bulk LSMO-032. A similar evolution from stripes to magnetic bubble domains can be expected in other bulk systems governed by a competition between ferromagnetic exchange and long-range dipole-dipole interaction, such as M-type hexaferrite,~\cite{yu-11} as well as in artificially fabricated multi-layered systems with opposing magnetic anisotropies, such as (Fe/Ni)/Cu/Ni/Cu(001).~\cite{wu-09} In the latter, the sandwiched out-of-plane magnetized Ni-layer competes with the in-plane magnetization of the top (Fe/Ni) layer via interlayer coupling. A transition between stripes and bubble domains can be achieved by varying the thickness of the separating Cu-layer. In contrast, LSMO-032 is located very close to an instability, which allows for an easy domain tuning via temperature and considerably low magnetic fields without changing the composition.

The composition of our specimen is also close to the one of the thin sample, in which a lattice of regularly arranged biskyrmions with $\approx 100$ nm diameter and the topological charge $N_{sk} = 2$ was reported (with $x = 0.315$).~\cite{yu-14} Although we cannot extract any information about the spin structure at the nanoscale from MFM, we classify our observed bubbles as topologically trivial magnetic bubbles with $N_{sk} = 0$. This assumption is supported by the fact that the bubbles are randomly arranged, appear to have varying diameters and shapes, and are significantly larger than the biskyrmions. Hence, the bubble formation is not (directly) related to the underlying crystallographic structure or to a delicate balance of competing interactions and anisotropies. On the other hand, the similarity in composition between the two samples suggests the possibility of observing a (bi-)skyrmion phase in our sample in the presence of an additional shape anisotropy. Indeed, Yu, et al., report a decisive dependence of the biskyrmion diameter on the sample thickness (i.e., on the shape anisotropy).~\cite{yu-14} This possibility opens the door of observing directly a crossover from topologically trivial magnetic bubble domains (in the bulk crystal) to topologically non-trivial skyrmions (in the thin sample limit) in the bilayer manganite LSMO-032 with a thickness gradient.

\section{Summary}

We presented a detailed investigation of the magnetic domain evolution in a LSMO-032 bulk sample in real space using MFM. Our field-dependent measurements uncovered the formation of bubbles from dendritic domains, with reversible domain structures but hysteretic magnetic magnitudes. In addition, our temperature dependent study shows the possibility of generating bubble domains through the field-cooling, as well as the gradual character of the SRT. Our results pave the way for future studies to reveal the relation between magnetic bubbles and skyrmions in materials with a high uniaxial anisotropy.

\begin{acknowledgments}
We gratefully acknowledge important discussions with N. Haberkorn. This work was supported by the Institute for Basic Science (IBS), Grant No. IBS-R014-D1, in Korea and NSF DMR 1122603 in the USA. Work at Argonne National Laboratory (crystal growth and sample characterization) was sponsored by the U.S. DOE, Office of Science, Basic Energy Science, Materials Science and Engineering Division. YHJ was supported by the Center for Topological Matter at POSTECH (2011-0030786).
\end{acknowledgments}

\end{document}